\documentclass[letter, 10pt, conference]{ieeeconf}      

\IEEEoverridecommandlockouts                              

\usepackage{graphics} 
\usepackage{epsfig} 
\usepackage{mathptmx} 
\usepackage{booktabs}
\usepackage{times} 
\usepackage{amsmath} 
\usepackage{amssymb}  
\usepackage{graphicx}
\usepackage{caption}
\usepackage{booktabs}
\usepackage[]{units} 
\usepackage{algorithm} 
\usepackage[noend]{algpseudocode} 
\algrenewcommand\algorithmicrequire{\textbf{Given: }} 
\DeclareMathAlphabet{\mathcal}{OMS}{cmsy}{m}{n} 
\usepackage{booktabs} 
\usepackage{graphicx}
\usepackage[caption=false]{subfig}
\usepackage{multirow}
\usepackage[table]{xcolor}
\usepackage{arydshln}
\usepackage{tikz}
\usetikzlibrary{calc}
\usepackage{bm}
\usepackage{algorithm,algpseudocode}
\makeatletter
\renewcommand*{\@opargbegintheorem}[3]{\trivlist
	\item[\hskip \labelsep{\bfseries #1\ #2}] \textit{(#3)}\ \itshape}
\makeatother

\graphicspath{ {Figures/} }
\usepackage{chngcntr}
\usepackage{pgfplots}	
\usepgfplotslibrary{external}
\definecolor{colori}  {rgb}{0.0 , 0.0 , 0.0 }
\definecolor{colorii} {rgb}{0.16, 0.31, 0.44}
\definecolor{coloriii}{rgb}{0.94, 0.59, 0.36}
\definecolor{coloriv} {rgb}{0.41, 0.54, 0.30}
\definecolor{colorv}  {rgb}{0.41, 0.16, 0.30}
\definecolor{halfgray}{rgb}{0.4 , 0.4 , 0.4 }
\definecolor{lightblue}{hsb}{0.55 , 0.85 , 0.79 }

\title{\LARGE \bf
Autoencoder-Based and Physically Motivated Koopman Lifted States for Wind Farm MPC: A Comparative Case Study
}

\author{Bindu Sharan$^{1}$, Antje Dittmer$^{2}$, Yongyuan Xu$^{1}$, and Herbert Werner$^{1}$ 
\thanks{$^{1}$ Bindu Sharan, Yongyuan Xu, and Herbert Werner are with Institute of Control Systems,
        Hamburg University of Technology, Germany.
        Emails: \{bindu.sharan, yongyuan.xu, h.werner\}@tuhh.de
	    }
\thanks{$^{2}$ Antje Dittmer is with the Institute of Flight Systems, German Aerospace Center, Email: antje.dittmer@dlr.de
  	    }
}

\pgfplotsset{compat=1.17}
\begin{document}
\maketitle
\thispagestyle{empty}
\pagestyle{empty}

\begin{abstract}
\noindent
This paper explores the use of Autoencoder (AE) models to identify Koopman-based linear representations for designing model predictive control (MPC) for wind farms. Wake interactions in wind farms are challenging to model, previously addressed with Koopman lifted states. In this study we investigate the performance of two AE models: The first AE model estimates the wind speeds acting on the turbines these are affected by changes in turbine control inputs. The wind speeds estimated by this AE model are then used in a second step to calculate the power output via a simple turbine model based on physical equations. The second AE model directly estimates the wind farm output, i.e., both turbine and wake dynamics are modeled. The primary inquiry of this study addresses whether any of these two AE-based models can surpass previously identified Koopman models based on physically motivated lifted states. We find that the first AE model, which estimates the wind speed and hence includes the wake dynamics, but excludes the turbine dynamics outperforms the existing physically motivated Koopman model. However, the second AE model, which estimates the farm power directly, underperforms when the turbines' underlying physical assumptions are correct.
We additionally investigate specific conditions under which the second, purely data-driven AE model can excel: Notably, when modeling assumptions, such as the wind turbine power coefficient, are erroneous and remain unchecked within the MPC controller. In such cases, the data-driven AE models, when updated with recent data reflecting changed system dynamics, can outperform physics-based models operating under outdated assumptions.
\end{abstract}
\section{Introduction}

Model Predictive Control (MPC) is a powerful technique for controlling nonlinear systems with time-varying parameters. However, its effectiveness relies on accurate models of the system dynamics, which can be challenging to obtain, especially for complex systems like wind farms, power grids, and autonomous vehicles. To address this challenge, researchers have turned to the Koopman operator \cite{koopman1931hamiltonian}, a mathematical tool that provides a linear representation of nonlinear dynamical systems. 
By using a linear Koopman model for nonlinear dynamics, nonlinear predictive control can be replaced by linear one, thus considerably reducing online computations and enabling real-time predictive control of systems that would otherwise be intractable.
The most important factor when using Koopman operators is the choice of lifted states, 
 the standard choices including orthogonal basis functions, e.g. radial basis functions and eigenfunctions of the Koopman operator. Also recently proposed was choosing lifted states based on physical insight i.e.,a priori knowledge of first principles model structure \cite{cisneros2020data,dittmer2023koopman,sharan2022real}.
In recent years, the integration of deep learning techniques with Koopman theory has shown promise in learning the dynamics of nonlinear systems. This approach leverages neural networks, particularly autoencoders (AE), to extract essential features from high-dimensional data, enabling the development of accurate system models. Notably, \cite{lusch2018deep} proposed modifying AE to learn Koopman operators.
Several studies have explored the application of deep Koopman models, among them \cite{ping2021deep}, where its application to transient stability in the power grid is shown. 
The system identification discussed in
 \cite{iacob2021deep} learns the linear parameter varying system representation of a nonlinear system 
 and \cite{shi2022deep} introduces an auxiliary control network to model state-dependent nonlinearity in control inputs.
%
Two types of cooperative wind farm control are wake redirection control (WRC), in which upstream turbines are yawed to deflect wakes away from downwind turbines, and axial induction control (AIC), which varies the thrust to change the axial induction factor.
%
In \cite{chen2022model} an MPC with combined AIC and WRC using deep learning-based reduced-order modelling is presented. The network architecture consists of a nonlinear encoder and decoder, and linear system and input matrices. An additional network predicts the power. This network is linearized to create output and throughput matrices.
The resulting model is used in a real-time capable MPC. However, the MPC relies on 700 longitudinal and lateral wind measurements distributed equidistantly over the two-dimensional area of a nine-turbine farm simulated in WFSim \cite{boersma2018control}. Similar to AIC in \cite{Cassamo.2020} and WRC in \cite{cassamo2021model}, which present a Koopman model-based linear MPC (KMPC) in the simulation tool SOFWA with a three-dimensional wind speed grid of hundred-thousand points, this number of simulated wind signals exceeds the number and granularity of wind measurements that can be reasonably expected to be available in wind farms. Using such a high number of measurement points might be both unrealistic and unnecessary: The quasi-linear MPC (qLMPC) presented in \cite{boersma2018control} which ships with WFSim shows good results assuming only the availability of wind speed measurements at the rotor disc. 
In \cite{sharan2022real} we designed an qLMPC estimating wind speeds without any measured wind for constant free-stream wind and in \cite{dittmer2022data} with only one additional wind speed measurement point for a step change in the free-stream wind. In \cite{dittmer2023koopman} we demonstrated a combined AIC-WRC KMPC to optimize power output.
A summary comparing the methodologies proposed in this study and previous studies of Koopman-based wind farm control approaches is presented in Table~\ref{tab:WFModels}.
\begin{table*}[ht]
    \begin{center}
        \caption {Comparison of windfarm MPC schemes in WFsim (related literature and proposed in this paper): Boersma et al., in dark blue, our previous work in light blue, Chen et al., in yellow and the proposed AE models in green. Control types are axial induction control (AIC, control by thrust $C_{T}^{’}$) and wake redirection control (WRC, control by yaw $\gamma$). Further differentiators are types of the free-stream $V_\infty$ and available signals:  turbine power $P_{Ti}$, wind in front of the i$^{th}$ turbine, $V_{i,m}$, effective wind speed $U_{ri}$ as the mean of perpendicular wind speeds at n rotor segments ($V_{i,1}, V_{i,2} \hdots V_{i,n}$) and $P_{\text{WF}}=\mathop {\sum}\limits_{i = 1}^{N_t} P_{Ti}$.}
        \label{tab:WFModels}
        \begin{tabular}{l l l l l l l l l} 
            \hline
            \hline
            Aspect & \cellcolor{blue!25}AIC\cite{boersma2018control} & \cellcolor{blue!10}AIC\cite{sharan2022real}& \cellcolor{blue!10}AIC\cite{dittmer2022data} & \cellcolor{blue!10}AIC+WRC\cite{dittmer2023koopman} & \cellcolor{yellow!25}AIC+WRC\cite{chen2022model} & \cellcolor{green!10}AIC (AE1) & \cellcolor{green!10}AIC (AE2) \\
            \hline
            $V_\infty$ speed & \cellcolor{blue!25}Constant &  \cellcolor{blue!10}Constant  & \cellcolor{blue!10}Changing  & \cellcolor{blue!10}Constant & \cellcolor{yellow!25}Constant& \cellcolor{green!10}Constant & \cellcolor{green!10}Constant\\
          
             Turbines & \cellcolor{blue!25} 9 &\cellcolor{blue!10}2  & \cellcolor{blue!10}2  & \cellcolor{blue!10}2 & \cellcolor{yellow!25}9 & \cellcolor{green!10}2 & \cellcolor{green!10}2\\
            \hline
            
            \textbf{Identification} & \cellcolor{blue!25}- & \cellcolor{blue!10}& \cellcolor{blue!10} &  \cellcolor{blue!10} & \cellcolor{yellow!25}&  \cellcolor{green!10} & \cellcolor{green!10}\\
            Lifted states & \cellcolor{blue!25}- &\cellcolor{blue!10}$6,12,24$ & \cellcolor{blue!10}$6$ &  \cellcolor{blue!10}$6$ & \cellcolor{yellow!25} $20$ &  \cellcolor{green!10} $20$ &\cellcolor{green!10} $2$ \\
            Input & \cellcolor{blue!25}- & \cellcolor{blue!10}$C_{T1}^{’}$, $C_{T2}^{’}$ &\cellcolor{blue!10}$C_{T1}^{’}$, $C_{T2}^{’}$, $V_{1,m}$  & \cellcolor{blue!10}$C_{T1}$,$C_{T2}$,$V_{1,m}$,$\gamma_1$ & \cellcolor{yellow!25}$C_{Ti}^{’}$, $\gamma_i$, where $i= 1,2 ...9$ &\cellcolor{green!10} $C_{T1}^{’}$, $C_{T2}^{’}$ &\cellcolor{green!10} $C_{T1}^{’}$, $C_{T2}^{’}$  \\
            Output & \cellcolor{blue!25}-  & \cellcolor{blue!10}$U_{r1}$,$U_{r2}$ & \cellcolor{blue!10}$U_{r1}$ ,$U_{r2}$  &  \cellcolor{blue!10}$P_{WF}$ & \cellcolor{yellow!25}$P_{WF}$ &\cellcolor{green!10}$U_{r1}$ ,$U_{r2}$ &\cellcolor{green!10} $P_{WF}$ \\
            \hdashline
            &\cellcolor{blue!25} & \cellcolor{blue!10}&\cellcolor{blue!10} & \cellcolor{blue!10} & \cellcolor{yellow!25} & \cellcolor{green!10} & \cellcolor{green!10} \\
            \multirow{-2}{5em}{Expert-Knowledge} & \multirow{-2}{*}{ \cellcolor{blue!25}- } &\cellcolor{blue!10}\multirow{-2}{4em}{required}  & \cellcolor{blue!10}\multirow{-2}{4em}{required} & \cellcolor{blue!10}\multirow{-2}{4em}{required} & \cellcolor{yellow!25}\multirow{-2}{*}{not required} & \cellcolor{green!10} \multirow{-2}{*}{not required} & \cellcolor{green!10} \multirow{-2}{*}{not required}   \\
            \hdashline
            System Id & \cellcolor{blue!25}-  &   \cellcolor{blue!10}EDMD & \cellcolor{blue!10}EDMD & \cellcolor{blue!10}EIODMD & \cellcolor{yellow!25} AE &  \cellcolor{green!10}AE+Koopman & \cellcolor{green!10} AE+Koopman  \\
            \hline
             &\cellcolor{blue!25} &\cellcolor{blue!10} & \cellcolor{blue!10}& \cellcolor{blue!10}& \cellcolor{yellow!25} &\cellcolor{green!10}& \cellcolor{green!10}\\
              &\cellcolor{blue!25} &\cellcolor{blue!10} & \cellcolor{blue!10}& \cellcolor{blue!10}& \cellcolor{yellow!25} &\cellcolor{green!10}& \cellcolor{green!10}\\
             \multirow{-3}{6em}{\textbf{Control} (No. of feedback signals)} &\multirow{-3}{6em}{\cellcolor{blue!25} 47 } &\multirow{-3}{6em}{\cellcolor{blue!10} 2} & \multirow{-3}{6em}{\cellcolor{blue!10} 3}& \multirow{-3}{6em}{\cellcolor{blue!10} 3}& \multirow{-3}{6em}{\cellcolor{yellow!25} 1400} &\multirow{-3}{6em}{\cellcolor{green!10} 2}& \multirow{-3}{6em}{\cellcolor{green!10} 2}\\
           
            \hdashline
             &\cellcolor{blue!25}- &\cellcolor{blue!10} & \cellcolor{blue!10}& \cellcolor{blue!10}& \cellcolor{yellow!25} &\cellcolor{green!10}& \cellcolor{green!10}\\
              &\cellcolor{blue!25}- &\cellcolor{blue!10} & \cellcolor{blue!10}& \cellcolor{blue!10}& \cellcolor{yellow!25} &\cellcolor{green!10}& \cellcolor{green!10}\\
           \multirow{-3}{6em}{Measurements} & \multirow{-3}{3em}{\cellcolor{blue!25}$P_{Ti}$, $U_{ri}$, $i= 1,2 \hdots 9$ }
                 & \multirow{-3}{6em}{\cellcolor{blue!10}$P_{T1}$,$P_{T2}$ } & \multirow{-3}{6em}{\cellcolor{blue!10}$P_{T1}$,$P_{T2}$, $V_{1,m}$ }  & \multirow{-3}{6em}{\cellcolor{blue!10}$P_{T1}$,$P_{T2}$, $V_{1,m}$ }& \multirow{-3}{6em}{\cellcolor{yellow!25}grid mesh $100\times 70$ } & \multirow{-3}{6em}{\cellcolor{green!10} $P_{T1}$,$P_{T2}$} & \multirow{-3}{6em}{\cellcolor{green!10} $P_{T1}$,$P_{T2}$}\\
               \hdashline
            Turbine Model & \cellcolor{blue!25}required &  \cellcolor{blue!10}required  & \cellcolor{blue!10}required  &  \cellcolor{blue!10}required  & \cellcolor{yellow!25}not required & \cellcolor{green!10} required & \cellcolor{green!10} not required \\
            \hdashline
            &\cellcolor{blue!25} &\cellcolor{blue!10} & \cellcolor{blue!10}& \cellcolor{blue!10}& \cellcolor{yellow!25} &\cellcolor{green!10}& \cellcolor{green!10}\\
              &\cellcolor{blue!25} &\cellcolor{blue!10} & \cellcolor{blue!10}& \cellcolor{blue!10}& \cellcolor{yellow!25} &\cellcolor{green!10}& \cellcolor{green!10}\\
             \multirow{-3}{6em}{Controller} &\multirow{-3}{3em}{\cellcolor{blue!25}NMPC} & \multirow{-3}{3em}{\cellcolor{blue!10}qLMPC} & \multirow{-3}{5em}{\cellcolor{blue!10}online- adaptive qLMPC} &  \multirow{-3}{3em}{\cellcolor{blue!10}KMPC}  & \multirow{-3}{3em}{\cellcolor{yellow!25}NMPC}& \multirow{-3}{6em}{\cellcolor{green!10} qLMPC} & \multirow{-3}{6em}{\cellcolor{green!10}KMPC} \\
             \hline
        \end{tabular}
    \end{center}
\end{table*}

The contribution of this paper is a case study that compares two different approaches to choosing Koopman lifted states AE-based vs. physically motivated on a benchmark example: predictive control of a wind farm, where a data-based Koopman operator is used to model the wake interaction between an upstream and a downstream turbine. 
AE used in the literature are typically undercomplete because they assume a large number of measured signals, i.e., messages or physical states. We showed in previous work that wake interaction models can be learned based on a small number of measured signals, and we propose to use AEs that are overcomplete: The dimension of code space is larger than the dimension of message space so that we can use the additional degrees of freedom to improve the model quality. Note that the lifted states do not have to be eigenfunctions, again using the additional freedom to improve the model.
 Models previously used in the wind farm control literature assume that wind speed measurements are available and can be used to predict the power output. In addition to the above, here we explore an alternative: utilising AE, to learn a model that predicts the power output directly without requiring wind speed measurements which will considerably simplify the implementation of the scheme. We demonstrate that this is possible at the expense of some performance degradation.

\section{Wind Farm Control} \label{sec:ProblemFormulation}
\begin{figure}\centering
\subfloat[Wind farm control]{\label{a}\includegraphics[width=0.48\textwidth,trim={0 1cm 1cm 0},clip]{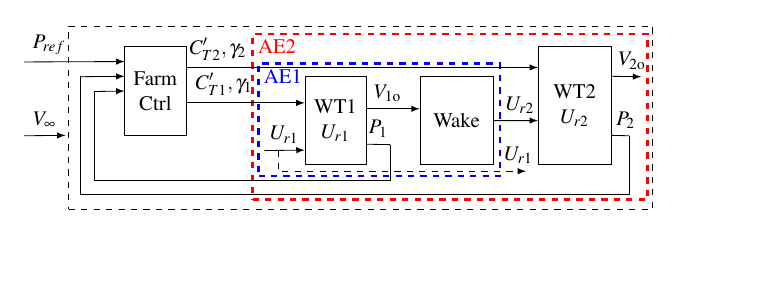}}\par
\subfloat[AE1 model]{\label{b}\includegraphics[width=.6\linewidth]{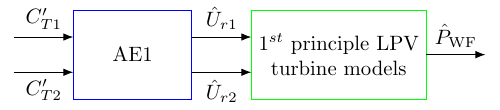}} 
\subfloat[AE2 model]{\label{c}\includegraphics[width=.35\linewidth]{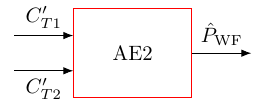}}
\caption{Wind farm control and identified models AE1 and AE2. Diagram (a) showcases the centralized farm-level controller incorporating the AE1 model for wind speed estimation (blue) and the AE2 model for power estimation (red). Diagram (b) illustrates the estimation of wind farm power through the integration of the AE1 model and first principle turbine models. Diagram (c) shows the direct estimation of farm power using the AE2 model.}
\label{fig:BlockDiagramFarm}
\end{figure}

Given a wind farm layout, a power grid reference $P_{ref}$, and free-stream wind $V_{\infty}$, turbine control inputs are to be set such that the power reference tracking error and the actuator activity are minimized.
In this work, we compare the performance of different centralized AIC MPC schemes to optimize the thrust control signals $C'_T$ for the wind farm layout depicted in Figure \ref{fig:BlockDiagramFarm}, with the AE1 model for wind speed estimation shown in blue and the AE2 model for power estimation in red. The wind farm consists of two NREL 5MW wind turbines with a 5D (5$\times$ the rotor diameter) distance. The free-stream is considered as constant with~8\,m/s, acting perpendicular on the first turbine. Hence, the yaw $\gamma$ is not changed in this work. For more details see the turbine and farm layout provided in \cite{sharan2022real}. 
The effective wind speed $U_{r\mathrm{i}}$ at the i$^{th}$ turbine is calculated as the mean speed at the $n_r$ rotor segments as
 \begin{align} 
 U_{r\text{i}}(\gamma_{\text{i}}) & = \mathrm{cos}(\gamma_{\text{i}}) \sqrt{\frac{1}{n_r} \sum_{j = 1}^{n_r} V_{lj}^2},  \quad  V_{lj}^2 = v_{xlj}^2 +v_{ylj}^2
 \end{align}
 with $v_x$ and $v_y$ as wind components in $x$ and $y$ directions respectively at coordinate point ($l$,$j$), the i$^{th}$ turbine's position in the simulated grid. For example, in our WFSim simulation, $U_{r1}$ is the root mean square of wind speeds $V_{11,11}$ to $V_{11,15}$ and the wind behind this turbine, denoted $V_{1\text{o}}$ in Figure \ref{fig:BlockDiagramFarm}, with energy extracted by the turbine, is the root mean square wind at the five grid points $V_{12,11}$ to $V_{12,15}$.
The discrete-time model for the i$^{th}$ turbine in the farm is:
        \begin{align} \label{eq:PowerWT}
            \begin{split}
             P_{\text{i},k+1} &  =  (1-\tau) \cdot P_{\text{i},k}+  \tau \cdot 0.5 \rho_a A_r (U_{r\text{i}}(\gamma_{\text{i},k}))^3 C_{T_\text{i},k}^{\prime}  
                \\
                \hat{C}_{T_{\text{i},k+1}}^\prime &=  (1-\tau)  \cdot\hat{C}_{T_\text{i},k}^\prime  + \tau \cdot C_{T_\text{i},k}^\prime\\
             \end{split}
        \end{align}
with  air density $\rho_a$, rotor area $A_r$, time constant $\tau$, state vector $x_{\text{WTi}} =[P_{\text{i}} \;\;\; \hat{C}_{T_\text{i}}^\prime]^T$, with $\hat{C}_{T_{\text{i}}}^\prime$ being the filtered thrust control. The control inputs are the thrust control signal $C_{T_\text{i}}^{\prime}$, as the yaw $\gamma_i$ controls are set to zero. Equation \eqref{eq:PowerWT} is  reformulated: 
\begin{align} \label{eq:WTWFsim}
  x_{\text{WTi},k+1} = A_{\text{WTi}} x_{\text{WTi},k} + B_{\text{WTi}}(U_{r\text{i}}(\gamma_i))C_{T_\text{i},k}^\prime, 
\end{align}
with the system and input matrices:
\begin{align*}
A_{\text{WTi}} = (1-\tau)I^{2\times2},\text{ } B_{\text{WTi}} = \tau[ C_P(U_{r\text{i}}(\gamma_i)), 1]^T.
\end{align*}
A concatenation into a wind farm model, with wind $U_r$ at and thrust $C_{T}^\prime$ and yaw $\gamma$ of all turbines, gives: 
\begin{align} \label{eq:LinWF}
       x_{\text{WF},k+1} = A_{\text{WF}} x_{\text{WF},k} +B_{\text{WF}}(U_r(\gamma)) C_{T,k}^\prime.
\end{align}
The resulting model is a quasi-linear parameter varying (qLPV) system, see \cite{hoffmann2014survey} for details on qLPV control. It is defined by the block diagonal system and input matrices $A_{\text{WF}} \in {\mathbb{R}}^{2n_T \times 2n_T}$ and $B_{\text{WF}} \in {\mathbb{R}}^{2n_T \times {n}_u}$ with turbine system and input matrices on their diagonals, $n_u$ control inputs for the farm, and farm state $x_{\text{WF}} \in {\mathbb{R}}^{2n_T}$.
 The effective wind  $U_{r2}$ is influenced by wakes, i.e., wind deficits at downwind turbines influenced both by free-stream and control action of upwind turbines.
 Wakes are causing slower wind speed, more turbulence, and wind direction changes \cite{boersma2017tutorial}.
Conventionally, wakes are modeled using the Navier-Stokes equations (NSE):
\begin{align*} 
 \frac{\partial \bm{v}}{\partial t}+(\bm{v}\cdot \nabla_H)\bm{v}+\nabla_H\cdot\bm{\tau}_h+\nabla_H p-\bm{f} = 0,\quad  
\nabla_H\cdot \bm{v} =  \frac{\partial v_y}{\partial y}
\end{align*}
with $\bm{v} =[v_x \;\; v_y]^T$, partial derivative $\nabla_H =[\partial/\partial x \;\; \partial/\partial y]^T$ and $p$ as the normalized pressure at hub height. 
$\bm{\tau}_h$ is the subgrid stress tensor in horizontal direction of the turbulence model and $\bm{f}$ denotes the effect of turbines on the flow. 
  
As no analytical solution exists, the NSE cannot be used to calculate effective wind in real-time. Measuring wind speeds at different grid points over a rotor disc can be done but requires expensive sensors, for example, spinner lidars \cite{mikkelsen2013spinner}. 

The baseline control scheme considered is our previously published Koopman-based qLMPC \cite{sharan2022real}, i.e., an MPC based on a qLPV wind farm model. In our previous work, Koopman models were used to estimate the effective wind speeds rather than assuming that the measurements are available. Th wind speeds are then used as scheduling parameter of the qLPV. The Koopman model has the same inputs and output as the AE1 model. To identify the Koopman models, known physical relationships of the NSEs were exploited when choosing the lifted states, i.e, moving average of wind speed over an extended period, and the difference between current wind speeds to include the temporal and spacial wake components, as well as the cubic relationship between power and $U_r$.

In this work, we want to investigate if we can get a similar or even better performance using no physical insight. Instead of designing Koopman lifted states via manually derived lifting function, lifted states are automatically set up using the integration of AE and Koopman framework.

We compare the baseline controller and the two AE-based controllers in two different scenarios: perfect knowledge of the turbine physics including the power coefficient and a divergence between real turbine and the turbine model used in MPC due to an unnoticed change in the power coefficient.

\section{Koopman-based Linear system identification}\label{sec:KoopmanIdentification}
The Koopman framework allows the representation of a nonlinear system with finite dimensions as a linear system with infinite dimensions. For detailed information, we refer to \cite{kaiser2020data} and \cite{proctor2018generalizing}. Additionally, extended dynamic mode decomposition (EDMD) as described in \cite{sharan2022real}, and extended input-output DMD (EIODMD), introduced in \cite{dittmer2023koopman}, are utilized to approximate the infinite-dimensional Koopman operator with a finite matrix. A brief overview of EDMD follows. Further elaboration and its application to partial differential equations can be found in \cite{arbabi2018data}. A discrete-time nonlinear dynamical system can be expressed as:
\begin{equation}
x_{k+1}=F(x_k,u_k)\label{eq:nonLinSys}
\end{equation}
with states $x\in\mathbb{R}^{n_{x}}$, inputs $u\in\mathbb{R}^{n_{u}}$, and nonlinear functions $F:\mathbb{R}^{n_{x}}\times \mathbb{R}^{n_{u}}\to \mathbb{R}^{n_{x}}$. 
Lifting functions $g:\mathbb{R}^{n_{x}}\to \mathbb{R}^{n_{g}}$ with $n_g > n_x$, are defined as nonlinear combinations of the original states $x$. 
A finite linear approximation of the nonlinear system is given in equation \eqref{eq:nonLinSys} as
\begin{align*}
g_{k+1}=Ag_{k}+Bu_{k}, \quad 
\hat{x}_k = Cg_k  ,\label{eq:linearPredictor}
\end{align*}
where $\hat{x}$ is the vector of predicted state $x$, $A\in \mathbb{R}^{n_g\times n_g}$ $B\in \mathbb{R}^{n_g\times n_u}$, $C\in \mathbb{R}^{n_x\times n_g}$. 
For EDMD, $n_o$ data samples containing measured state and input vectors are collected as:
\begin{align*}
\mathcal{D}=\left\{ \left[
x_{k}^T,u_{k}^T
\right]^T\right\} _{k=1}^{n_{O}-1},
\end{align*}
and matrices $A,B$ and $C$ are obtained via the optimization: 
\begin{equation}
\min_{{K}}\sum_{k=1}^{n_{O}-1}\left\Vert 
g(x_{k+1})
-{K}\begin{bmatrix}
g(x_k)\\u_k
\end{bmatrix}\right\Vert _{2}^{2}\label{eq:ApproxKalman}
\end{equation}
\begin{equation}
\min_{{C}}\sum_{k=1}^{n_{O}-1}\left\Vert 
x_{k}-Cg(x_k)
\right\Vert _{2}^{2}\label{eq:ApproxC}
\end{equation}
with the Koopman matrix
$
{K}=\begin{bmatrix}A & B \end{bmatrix} \in \mathbb{R}^{n_g\times (n_g + n_u)}.
$
The lifting functions are applied to the data set $\mathcal{D}$
to design the following matrices
\begin{align*}
 G_u &=\begin{bmatrix}{g}(x_1) & \cdots& g(x_{n_{O}-1})\\
    u_{1} &\cdots& u_{n_{O}-1} \end{bmatrix} \in \mathbb{R}^{(n_g +n_u)\times (n_0-1)}\\
 G_{+} &=\begin{bmatrix}g(x_{2}) &\cdots& g(x_{n_{O}})
\end{bmatrix} \in \mathbb{R}^{n_g \times (n_0-1)}\\
X_{+} &=\begin{bmatrix}x_{2} &\cdots& x_{n_{O}}
\end{bmatrix} \in \mathbb{R}^{n \times (n_0-1)}
\end{align*}
which are used in a reformulation of the optimization problems \eqref{eq:ApproxKalman} and \eqref{eq:ApproxC} as
\[
\min_{{K}}\left\Vert G_{+}-{K}G_u\right\Vert _{F}^{2},\quad \min_{{C}}\left\Vert X_{+}-CG_{+}\right\Vert _{F}^{2}
\] 
where $\left\Vert. \right\Vert_{F}$ denotes the Frobenius norm. The analytical solution
to this linear least square problem is obtained as 
\begin{align*}
{K}= G_{+}G_u^{\dagger}, \quad C= X_{+}G_+^{\dagger}
\end{align*}
where $^{\dagger}$ denotes the Moore-Penrose pseudoinverse. 
To obtain the Koopman-based linear model for wind farms, different set of observables motivated by NSE have been considered and explained in \cite{sharan2022real}. 
\subsection{ AE for Identification of the Koopman Matrix}
\begin{figure}
    \centering
    \includegraphics[width=0.48\textwidth]{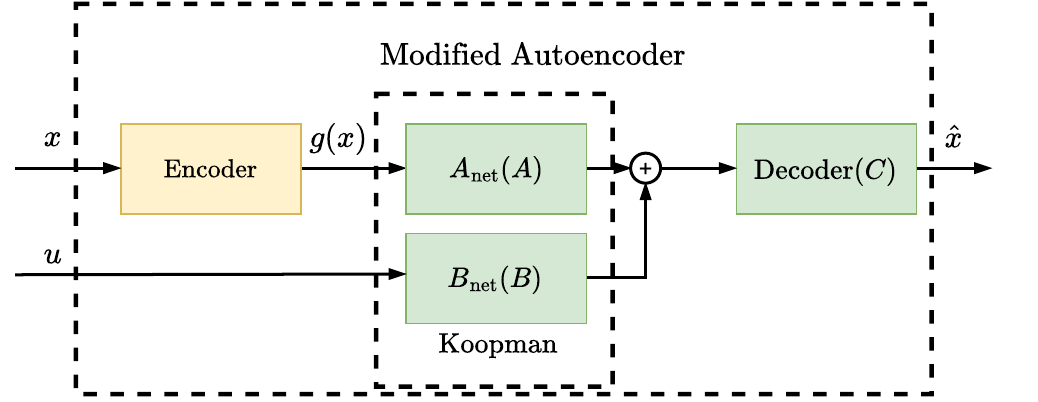}
    \caption{Identification of linear system representation of a nonlinear system by integrating Koopman and AE}
    \label{fig:ModiAE}
\end{figure}

In the previous subsection, we observed that identifying the Koopman matrix requires the selection of lifted states a priori. Literature suggests that leveraging physical insight aids in choosing these states instead of opting for random nonlinear functions, e.g., monomials, and radial basis functions, see \cite{cisneros2020data,dittmer2023koopman,sharan2022real}. In this section, AE will be employed to identify the Koopman model and lifted states simultaneously without any physical insights. It will be used not only to minimize one-step prediction error by optimizing the cost functions \eqref{eq:ApproxKalman} and \eqref{eq:ApproxC}, but also to minimize multi-step prediction error by minimizing the loss function outlined in \cite{lusch2018deep}. 
AE mainly consists of an encoder and a decoder with the same level of complexity in both. However, to learn the lifted states and Koopman operator simultaneously, AE is modified by adding an auxiliary network in between the encoder and decoder as shown in figure~\ref{fig:ModiAE}. The yellow colour and green colour in figure~\ref{fig:ModiAE}, signify the nonlinear and linear structure of networks respectively. In this study, we identify a Koopman model together with lifted states by minimizing the subsequent cost function:
\begin{equation}\label{eq:loss_function}
 {\cal L} =  \alpha _1{\cal L}_{\mathrm{recon}} + \alpha _2{\cal L}_{\mathrm{pred}} + \alpha _3{\cal L}_{\mathrm{lin}}.  
\end{equation}
where $\alpha_i$ are weighting factors. ${\cal L}_{\mathrm{recon}}$, ${\cal L}_{\mathrm{pred}}$ and ${\cal L}_{\mathrm{lin}}$ are calculated as:
\begin{eqnarray*}
{\cal L}_{\mathrm{recon}} &=&  \frac{1}{{n_o - 1}}\mathop {\sum}\limits_{i = 1}^{n_o - 1}\left\| {x}_i - \mathrm{dec}(\mathrm{enc} (x_i)) \right\|_{\mathrm{MSE}}\\
&=& \frac{1}{{n_o - 1}}\mathop {\sum}\limits_{i = 1}^{n_o - 1}\left\| {x}_i - C(g (x_i)) \right\|_{\mathrm{MSE}}\\
{\cal L}_{\mathrm{pred}}&=& \frac{1}{{n_o - 1}}\mathop {\sum}\limits_{k = 1}^{n_o - 1} \bigg[\frac{1}{{N_p}}\mathop {\sum}\limits_{i = 1}^{N_p} \left\| {x}_{i+k } - \hat{x}_{i+k} \right\|_{\mathrm{MSE}}\bigg]\\
{\cal L}_{\mathrm{lin}}&=& \frac{1}{{n_o - 1}}\mathop {\sum}\limits_{k = 1}^{n_o - 1} \bigg[\frac{1}{{N_p}}\mathop {\sum}\limits_{i = 1}^{N_p} \left\|g( {x}_{i+k }) - \hat{g}({i+k} )\right\|_{\mathrm{MSE}}\bigg]\\
\end{eqnarray*}
where $\hat{x}_{i+k} = \mathrm{dec}(\hat{g}_{k+i})$, $\hat{g}(k+i) = A^{i}g(x_{k})+ \mathcal{B}_i \mathcal{U}_i\implies 
\hat{x}_{i+k} = CA^{i}g(x_{k})+ C \mathcal{B}_i \mathcal{U}_i \text{, } \mathcal{B}_i = \begin{bmatrix} A^{i-1}B & A^{i-2}B & \hdots & B    
\end{bmatrix}$, and \\$\mathcal{U}_i=\begin{bmatrix} u(k)^T& u(k+1)^T& \hdots & u(k+i-1)^T   
\end{bmatrix}^T.$

We employed the AE to derive $A$, $B$, and $C$ matrices, with the linear $C$ enabling real-time MPC by reformulation as a quadratic programming (QP) problem. The training of the network can be executed through two primary methods, single-level training and bi-level training. These approaches play a crucial role in training efficiency see \cite{huang2023learning}. More details about these training methods can be found in Appendix \ref{Appendix: training}.

\section{AE-based Wind Farm Model} \label{secWakeModel} 
To design a wind farm controller, both turbine and wake models are needed, with the latter ones often implicitly used by leveraging wind measurements. In the literature (see Table~\ref{tab:WFModels}), either a high number of wind measurements are required or physical first principles are used. In previous works, we demonstrated a considerable reduction of the number of wind measurements by using physically motivated lifted states for K-qLMPC. This paper investigates the usage of AE models for wind farm MPC without physical insight and without any wind measurements. Below, the architectures of model AE1, for wind estimation including wake effects, and model AE2, for farm power estimation including wake and turbine dynamics, are described.  Both models are trained on simulated open-loop data from WFSim without any physical information provided in the training process.



\subsection{AE-based Koopman Model for Effective Wind Estimation}
The AE-based wake model uses the same measurements as our previous work \cite{sharan2022real}. The network architecture is depicted in Figure \ref{fig:BlockDiagramNNArch1}, with single-level training outlined in Algorithm~\ref{alg:Single level}. Four models with different numbers of lifted states $N_\phi$ were designed, used in MPC and compared regarding their performance. The encoder has a sigmoid activation function and 3 hidden layers, each with $n_e$ activation functions, while the decoder includes 3 hidden layers with size $N_g\times2n_e\times2n_e$. The network reprenting matrices $A$ and $B$ have one hidden layer with $N_g$ activation functions. We set $\alpha_1 = \alpha_2 = 0.45$ and $\alpha_3 = 0.1$, with 500 epochs and a batch size of 50. The model with 6 lifted states has $n_e = 124$ and the models with 12, 24 and 50 lifted states have $n_e = 496$ activation functions.
 \subsection{AE-based Koopman Model for Farm Power Estimation}

\begin{figure}[ht]
\centering
  \includegraphics[width=0.24\textwidth]{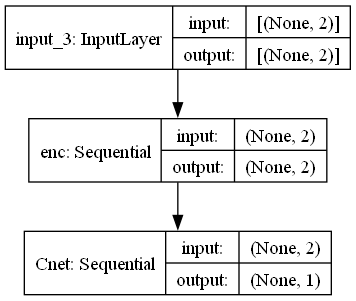}
    \caption{Block diagram NN Architecture for bi-level training}
    \label{fig:BlockDiagramNNArch2}
\end{figure}

The AE-based wind farm model incorporates the power of both turbines as physical states. Two models are trained, with 2 and 6 lifted states respectively. The architecture is identical, but for the 2, respectively 6, input layers in the decoder. The network (Figure \ref{fig:BlockDiagramNNArch2}) employs bi-level training (Algorithm \ref{alg:Bi-level}) and uses swish nonlinear activation functions for the encoder. The encoder comprises 3 hidden layers with 20 activation functions each and the decoder (\verb|Cnet|) uses linear activation functions and an output layer with 1 activation function. Matrices $A$ and $B$ are obtained using EDMD in each epoch. We set $\alpha_1 = \alpha_2 = \alpha_3 = 1$ with 500 epochs and a prediction horizon of $N_p = 10$. 


\section{Model Predictive control} \label{sec:qLMPCDesign}
In the previous section, we identified two different models using AE and Koopman theory: An AE-based Koopman model for effective wind speed estimation named AE1, and one for wind farm power, which we refer to as the AE2 model. 
In the subsections below, descriptions of qLMPC based on AE1 and of KMPC, which uses AE2, are provided. 
 \subsection{qLMPC Design using AE1 model}
The aforementioned objectives of wind farm MPC, small power reference tracking errors and minimal changes in control inputs, can be formulated in a cost function 
 \begin{align} \label{eq:NMPCcost}
     \begin{array}{l} 
         \underset{\bm{U}_k}{\text{min }} J(\bm{U}_k) = E_k^T Q E_k +\Delta \bm{U}_k^T R \Delta \bm{U}_k\\
         s.t.\\
         \begin{bmatrix} x_{\text{WF},k+1} \end{bmatrix} = \begin{bmatrix} A_{\text{WF}} & B_{\text{WF}} (\rho) \end{bmatrix} \begin{bmatrix} x_{\text{WF},k}\\ U_{k}\end{bmatrix},\\ \quad   k \in \{1, 2, \dots, n_h-1\}, \\
     \end{array}
 \end{align}
 based on the trajectories of farm power tracking error $E_k$ 
 \begin{align*}
E_k = [ e_k,e_{k+1},\ldotp \ldotp \ldotp ,e_{k+n_h-1} ]^T \in \mathbb{R}^{n_h\times 1} \end{align*}
where the power reference tracking error in time step $k$ on farm level is calculated as $e_k = P_{re,k} - P_{\text{WF},k}.$ 
 The change in control inputs $\Delta \bm{U}_k \in \mathbb{R}^{n_h n_u \times 1}$ is calculated as
\begin{align*}
\left\lbrack U_k^T -U_{k-1}^T,U_{k+1}^T -U_k^T ,\ldotp \ldotp \ldotp, U^T_{k+n_h-1} - U^T_{k+n_h-2} \right\rbrack^T
\end{align*}
 with the vector $U_k \in \mathbb{R}^{n_u \times 1}$ containing all control inputs for all turbines at time step $k$.
 The weighting matrices are based on the scalar weights $q$ and $r$ as $Q = q \cdot I^{{n_h}\times{n_h}}$ and $R = r \cdot I^{{n_u n_h}\times{n_u n_h}}$ with $n_h$ sample steps of the preceding horizon and $n_u$ the number of control inputs. The system and input matrices $A_{\text{WF}}$ and $B_{\text{WF}}$ are as defined in equation \eqref{eq:LinWF}, as is the state vector $x_{\text{WF}}$. The system is scheduled on the effective wind speeds with $\rho = [ U_{r1}, U_{r2}]$. 
\\
  For both AE designs, the cost function of equation \eqref{eq:NMPCcost} can be written for the quadratic programming (QP) formulation as 
 \begin{align*} 
 J(\bm{U}_k) &= (\tilde{L} x_0+\tilde{S}\bm{U}_k -P_{ref})^T Q(\tilde{L} x_0+\tilde{S} \bm{U}_k - P_{ref}) \\
 &+ \Delta \bm{U}_k^T  R \Delta\bm{U}_k
\end{align*}
For the AE1 the initial state $x_0 \in {\mathbb{R}}^{2n_T}$ is the state from the last sample and the matrices $\tilde{L}_{\text{AE1}}\in\mathbb{R}^{n_h\times2n_T}$ and $\tilde{S}_{\text{AE1}}\in\mathbb{R}^{n_h\times n_h n_u}$, which calculate the expected future trajectory of the farm power $P_T$. They are derived from the Toeplitz matrices $\Lambda_{\text{AE1}}$ and $S_{\text{AE1}}$, see \cite{sharan2022real}, which give the expected future states for the preceding horizon trajectory. 
%
 \subsection{KMPC Design using AE2 model}
The power $P_{\text{WF}}$ is estimated directly in the approach using the 
 AE2 model. The result is a linear MPC as this omits the effective wind speeds as scheduling parameters. For this second control algorithm, the stacked Toeplitz matrices $\Lambda_{\text{AE2}}$ and $S_{\text{AE2}}$ are based on the input, system and output matrices $A_{\text{AE2}}$, $B_\text{AE2}$, and $C_{\text{AE2}}$ as
 {
\begin{align*}
    \Lambda_{\text{AE2}} &=\\ & \begin{bmatrix}
(C_{\text{AE2}} A_{\text{AE2}})^T & (C_{\text{AE2}} A_{\text{AE2}}^{2})^T & \cdots  &(C_{\text{AE2}} A_{\text{AE2}}^{n_h-1})^T
\end{bmatrix}^T, \\
S_{\text{AE2}} &= \\
&\begin{bmatrix}
C_{\text{AE2}} B_{\text{AE2}} &  0 & \cdots &0\\
\vdots & \vdots &\cdots &0 \\
C_{\text{AE2}} A_{\text{AE2}} ^{n_h-2} B_{\text{AE2}} & C_{\text{AE2}}A_{\text{AE2}}^{n_h-3}B_{\text{AE2}}& ...&C_{\text{AE2}} B_{\text{AE2}}\\
\end{bmatrix}.
\end{align*}}%
Note that, unlike for the AE1 model, for the AE2 model which estimates $P_{\text{WF}}$ there is $\tilde{L}_{\text{AE2}} = \Lambda_{\text{AE2}}$ and $\tilde{S}_{\text{AE2}} = S_{\text{AE2}}$.\\
The next section presents system identification results obtained for AE1 and AE2 and closed-loop simulation tracking performance for the corresponding K-qLMPC and KMPC. 

\section{Simulation Results} \label{sec:SimResults}
The performance of the AE-based models with different numbers of lifted states $N_g$, as described in Section \ref{sec:KoopmanIdentification}, is evaluated regarding their variance-accounted-for (VAF) using a dataset obtained in an open-loop WFSim simulation with the thrust control signals set to band-limited white noise. The data is split into a training and a validation part, see Table II for the VAF of the validation data. The AE1 model with $N_g = 6$ performs best and an increase in number of lifted states does not improve the VAF. However, a greater VAF does not necessarily correlate to a smaller tracking error (TE), as the occurrence of frequencies excited in closed loop is influenced by the controller. Hence, the closed loop tracking error of the corresponding MPC designs was evaluated for all four trained AE1 and the two AE2 models. For the AE1 model, it was noted that while the VAF decreases slightly with increasing $N_g$, the TE of the AE1 models with 24 states is considerably smaller than the AE1 model with 6 and 12 $N_g$. As an increase from $N_g = 24$ to $N_g = 50$ did not result in further improvement, but indeed yielded a slight increase of TE on the selected power reference, the AE1 model with $N_g = 24$ was selected. For AE2 an increase from 2 to 6 $N_g$ did not show any changes, neither in VAF nor in TE. Hence, the simpler model with 2 lifted states is selected. Due to lack of time, no further increase of $N_g$ was tested. 

\begin{table}[ht]
\begin{center}
\caption{Identification and Closed-Loop Performance Comparison of Various Autoencoder (AE) Models with number of lifted states~$N_g$, effective wind speed of the i$^{th}$ turbine~$U_{ri}$, farm power~$P_{\text{WF}}$, and tracking error TE}
\setlength{\tabcolsep}{3.9pt}\label{tab:Identification1}
\begin{tabular}{lrrrrrrrr} 
\hline
\hline
   \multirow{2}{2em}{Model} &  \multirow{2}{4em}{Training parameters} & \multirow{2}{*}{$N_g$} & \multicolumn{3}{c}{VAF [\%]}&  \multirow{2}{*}{Controller}&  \multirow{2}{2em}{TE\,[kW]}\\
  & & & $U_{r1}$ & $U_{r2}$ & $P_{\text{WF}}$ &\\
  \hline
\multirow{4}{*}{AE1} & 79,976   & 6 & 97 & 87 & - &qlMPC &55.7 \\

                    & 1,251,368 & 12 & 93  & 74  & -&qlMPC  &45.7 \\
                    & 1,270,544 & 24 & 92  & 78  & - &qlMPC &17.3 \\
                    &1,315,056  & 50 & 93  & 73  & - &qlMPC &18.1 \\
  \hline
  \multirow{3}{*}{AE2} & 882 & 2 & - & - & 52 &KMPC & 66.9 \\
                      & 966 & 6  & - & - &  52 &KMPC  & 66.8 \\
\hline
\hline
\end{tabular}
\end{center}

 \label{tab:TE}
\end{table}
The qLMPC and KMPC controllers' performance, utilizing  AE1 and AE2 models respectively, is assessed based on a reference signal $P_{\text{ref}}(k)$ defined as follows:
\[
   P_{ref}(k)= 
\begin{cases}
   0.8\cdot P_{greeedy}+0.35 \cdot P_{greeedy}\cdot\delta P(k),& \text{if } k\leq 400\\
    0.95\cdot P_{greeedy}+0.15 \cdot P_{greeedy}\cdot\delta P(k),              & \text{otherwise}
\end{cases}
\]
Here, $P_{greedy}$ represents the total farm power generated by operating both turbines with maximum thrust coefficient and aligned with the wind. The additional power $\delta P(k)$ simulates grid power demands and is taken from \cite{boersma2019constrained}. 

Subsequently, the closed-loop performance of three MPC schemes is compared: our previous K\,qLMPC design with~24 physically motivated lifting function \cite{sharan2022real}, denoted as K$_{24}$,qLMPC, and the two AE model based MPC designs, K$_{\text{AE1}}$ and K$_{\text{AE2}}$. They are compared under two scenarios: 
 \begin{itemize}
    \item \textbf{Scenario 1:} Identical MPC model and system, including identical power coefficients ($c_p$)
    \item  \textbf{Scenario 2:} Nonidentical MPC model and system with $0.003\%$ offset change in power coefficient ($c_p$) due to aging
\end{itemize}
As before, the performance is quantified using the tracking error (TE) and the actuator activity (AA) with the MPC cost function from Equation \eqref{eq:NMPCcost} with weighing matrices $Q$ and $R$ set to $10^{-4}$ and $10^{-6}$ respectively. 

\begin{figure}[ht]
  \includegraphics[width= 0.48\textwidth]{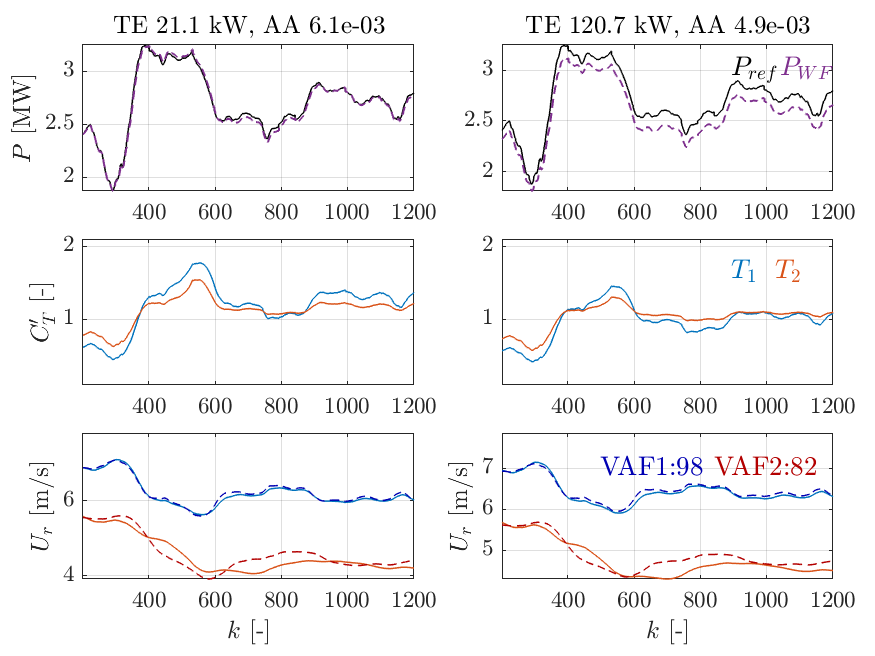}
    \caption{Farm power reference tracking (K$_{24}$\,qLMPC). Column 1 corresponds to scenarios 1 and column 2 corresponds to scenarios 2 }
    \label{fig:K24}
\end{figure}

\begin{figure}[ht]
 \includegraphics[width=0.48\textwidth]{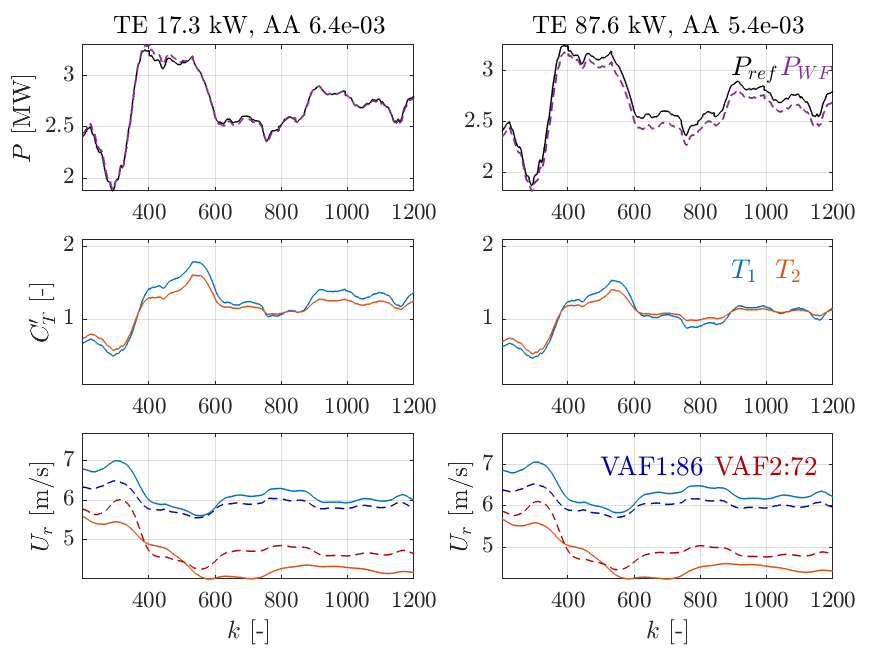}
    \caption{Farm power reference tracking (K$_{\text{AE1(24)}}$\,qLMPC), Column 1 corresponds to scenarios 1 and column 2 corresponds to scenarios 2 }
    \label{fig:AE1}
\end{figure}

\begin{figure}[ht]
 \includegraphics[width=0.48\textwidth]{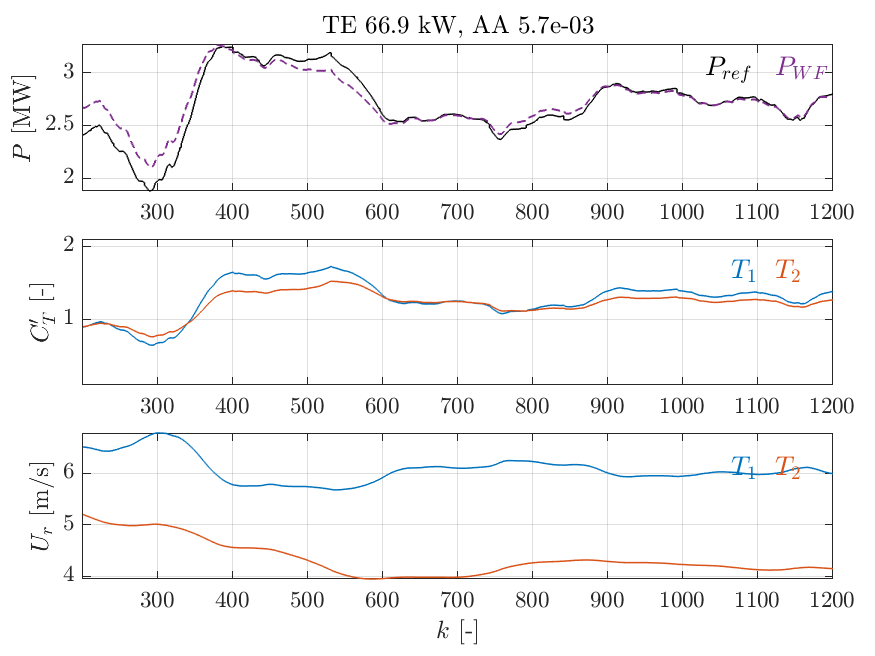}
    \caption{Farm power reference tracking (K$_{\text{AE2}}$\,qLMPC), in both scenarios}
    \label{fig:AE2}
\end{figure}

Simulation results are summarized in Table~\ref{tab:compcont}  and depicted in Figures \ref{fig:K24}, \ref{fig:AE1} and \ref{fig:AE2}. In all figures, the reference wind farm power $P_{\text{ref}}$ is represented by a solid black line, while the actual farm power $P_{\text{WF}}$ is depicted by a dashed purple line in the first row of each subplot. The second and third rows display the control action and effective wind velocities, respectively. 
Figures \ref{fig:K24} and \ref{fig:AE1} depict the closed-loop performance with K$_{24}$\,qLMPC and K$_{\text{AE1}}$\,qLMPC for the two scenarios. We see a slight improvement for AE1 in both scenarios, demonstrating that AE are indeed able to learn the wake dynamics. However, both qLMPC controllers' performance deteriorates considerably in Scenario 2 due to the reliance on the turbine models.
Figure \ref{fig:AE2} presents the closed-loop performance of K$_{\text{AE2}}$\,MPC which was designed to mitigate the effects of changing $C_p$ and reduce the number of measurements. As this KMPC does not use $c_P$ explicitly the simulation results are the same for both scenarios. 
Notably, K$_{\text{AE2}}$\,MPC exhibits inferior performance in Scenario~1. This is due to the direct identification of the wind farm power, without relying on the effective wind speed, as well as the absence of a physical-equation-based turbine model. However, in Scenario 2, K$_{\text{AE2}}$\,MPC outperforms the qLMPC controllers by a factor of 2.
Table~\ref{tab:compcont} 
summarizes the simulation results, highlighting the superior performance of the K$_{\text{AE1}}$ controller for a complete match of MPC model and actual turbine (Scenario 1), whereas K$_{\text{AE2}}$\,MPC outperforms in Scenario 2  with a $c_p$ mismatch.

In summary, we were able to replace the previously suggested Koopman model for wakes based on physically motivated lifted states with an AE Koopman model that does not rely on a priori knowledge. However, an AE model replacing both the wake and turbine model did result in some performance degradation. We could show the possible advantages of using such an AE model for farm power prediction by simulating a discrepancy between real and modelled turbines by changing the power coefficient.
These findings suggest avenues for future research, combining data-driven AE wind farm models with physical insights, e.g., by incorporating inherent delays by using Long Short-Term Memory units (LSTM). Once an AE model architecture is designed, the system and input matrices  
can be used to online update the wind farm model using the approach provided in \cite{dittmer2022data} to make this model robust against changes, e.g., in the power coefficient due to rotor blade degradation.

\begin{table}[ht]
\begin{center}
\caption{Controller comparison in scenario 1 (identical MPC model and system) and scenario 2 (nonidentical MPC model and system due to deteriorated power coefficient) with regard to tracking error TE and actuator activity AA }
\setlength{\tabcolsep}{3.9pt}\label{tab:compcont}
\begin{tabular}{lrrrrrrrrrr} 
\hline
\hline
 & \multicolumn{2}{c}{K$_{24}$\,qLMPC} & \multicolumn{2}{c}{K$_{\text{AE1}}$\,qLMPC} & \multicolumn{1}{c}{K$_{\text{AE2}}$\,MPC }\\
Lifted states & \multicolumn{2}{c}{24}&  \multicolumn{2}{c}{24}& \multicolumn{2}{c}{2}\\
Scenario & 1& 2& 1& 2 & 1, 2\\
\hline
TE\,[kW] & 21.6  & 120.7 &  17.3 & 87.6 & 66.9\\
AA/10$^-3$& 6.1   & 4.9  &   6.4 & 5.4   & 5.7\\
\hline
\hline
\end{tabular}
\end{center}
\end{table}

\section{Conclusion} \label{sec:Conclusion}
In this study, AE and Koopman theory were combined to identify models for wind farm MPC. A streamlined bi-level training enables representing nonlinear dynamics as a Koopman-based linear model. 
We show that an AE based MPC design, with the AE model estimating the wind speeds, outperforms the previously published MPC based on Koopman models using physically motivated lifting functions.  
We conclude that AE models estimating the wind farm power without any assumptions regarding the relationship between wind and power and without using any wind measurements have the potential to make the controller more robust against modelling assumption deficiencies.
Future research could explore various Recurrent Neural Network structures that offer to model the temporal and spatial dynamics of wind farms for variable free wind speed and direction: such as Gated Recurrent Units (GRU), Long Short-Term Memory units (LSTM), and temporal convolutional networks.






\bibliographystyle{ieeetr}
\bibliography{myLiterature.bib}

\begin{thebibliography}{10}

\bibitem{koopman1931hamiltonian}
B.~O. Koopman, ``Hamiltonian systems and transformation in hilbert space,''
  {\em Proceedings of the national academy of sciences of the united states of
  america}, vol.~17, no.~5, p.~315, 1931.

\bibitem{cisneros2020data}
P.~S. Cisneros, A.~Datar, P.~G{\"o}ttsch, and H.~Werner, ``Data-driven
  quasi-lpv model predictive control using koopman operator techniques,'' {\em
  IFAC-PapersOnLine}, vol.~53, no.~2, pp.~6062--6068, 2020.

\bibitem{dittmer2023koopman}
A.~Dittmer, B.~Sharan, and H.~Werner, ``Koopman model predictive control for
  wind farm yield optimization with combined thrust and yaw control,'' {\em
  IFAC-PapersOnLine}, vol.~56, no.~2, pp.~8420--8425, 2023.

\bibitem{sharan2022real}
B.~Sharan, A.~Dittmer, and H.~Werner, ``Real-time model predictive control for
  wind farms: a koopman dynamic mode decomposition approach,'' in {\em 2022
  European Control Conference (ECC)}, pp.~1006--1011, IEEE, 2022.

\bibitem{lusch2018deep}
B.~Lusch, J.~N. Kutz, and S.~L. Brunton, ``Deep learning for universal linear
  embeddings of nonlinear dynamics,'' {\em Nature Communications}, vol.~9,
  no.~1, p.~4950, 2018.

\bibitem{ping2021deep}
Z.~Ping, Z.~Yin, X.~Li, Y.~Liu, and T.~Yang, ``Deep koopman model predictive
  control for enhancing transient stability in power grids,'' {\em
  International Journal of Robust and Nonlinear Control}, vol.~31, no.~6,
  pp.~1964--1978, 2021.

\bibitem{iacob2021deep}
L.~C. Iacob, G.~I. Beintema, M.~Schoukens, and R.~T{\'o}th, ``Deep
  identification of nonlinear systems in koopman form,'' in {\em 2021 60th IEEE
  Conference on Decision and Control (CDC)}, pp.~2288--2293, IEEE, 2021.

\bibitem{shi2022deep}
H.~Shi and M.~Q.-H. Meng, ``Deep koopman operator with control for nonlinear
  systems,'' {\em IEEE Robotics and Automation Letters}, vol.~7, no.~3,
  pp.~7700--7707, 2022.

\bibitem{chen2022model}
K.~Chen, J.~Lin, Y.~Qiu, F.~Liu, and Y.~Song, ``Model predictive control for
  wind farm power tracking with deep learning-based reduced order modeling,''
  {\em IEEE Transactions on Industrial Informatics}, vol.~18, no.~11,
  pp.~7484--7493, 2022.

\bibitem{boersma2018control}
S.~Boersma, B.~Doekemeijer, M.~Vali, J.~Meyers, and J.-W. van Wingerden, ``A
  control-oriented dynamic wind farm model: Wfsim,'' {\em Wind Energy Science},
  vol.~3, no.~1, pp.~75--95, 2018.

\bibitem{Cassamo.2020}
N.~Cassamo and J.-W. {van Wingerden}, ``On the potential of reduced order
  models for wind farm control: A koopman dynamic mode decomposition
  approach,'' {\em Energies}, vol.~13, no.~24, p.~6513, 2020.

\bibitem{cassamo2021model}
N.~Cassamo and J.-W. van Wingerden, ``Model predictive control for wake
  redirection in wind farms: a koopman dynamic mode decomposition approach,''
  in {\em 2021 American Control Conference (ACC)}, pp.~1776--1782, IEEE, 2021.

\bibitem{dittmer2022data}
A.~Dittmer, B.~Sharan, and H.~Werner, ``Data-driven adaptive model predictive
  control for wind farms: A koopman-based online learning approach,'' in {\em
  2022 IEEE 61st Conference on Decision and Control (CDC)}, pp.~1999--2004,
  IEEE, 2022.

\bibitem{hoffmann2014survey}
C.~Hoffmann and H.~Werner, ``A survey of linear parameter-varying control
  applications validated by experiments or high-fidelity simulations,'' {\em
  IEEE Transactions on Control Systems Technology}, vol.~23, no.~2,
  pp.~416--433, 2014.

\bibitem{boersma2017tutorial}
S.~Boersma, B.~M. Doekemeijer, P.~M. Gebraad, P.~A. Fleming, J.~Annoni, A.~K.
  Scholbrock, J.~A. Frederik, and J.-W. van Wingerden, ``A tutorial on
  control-oriented modeling and control of wind farms,'' in {\em 2017 American
  control conference (ACC)}, pp.~1--18, IEEE, 2017.

\bibitem{mikkelsen2013spinner}
T.~Mikkelsen, N.~Angelou, K.~Hansen, M.~Sj{\"o}holm, M.~Harris, C.~Slinger,
  P.~Hadley, R.~Scullion, G.~Ellis, and G.~Vives, ``A spinner-integrated wind
  lidar for enhanced wind turbine control,'' {\em Wind Energy}, vol.~16, no.~4,
  pp.~625--643, 2013.

\bibitem{kaiser2020data}
E.~Kaiser, J.~N. Kutz, and S.~L. Brunton, ``Data-driven approximations of
  dynamical systems operators for control,'' in {\em The Koopman Operator in
  Systems and Control}, pp.~197--234, Springer, 2020.

\bibitem{proctor2018generalizing}
J.~L. Proctor, S.~L. Brunton, and J.~N. Kutz, ``Generalizing koopman theory to
  allow for inputs and control,'' {\em SIAM Journal on Applied Dynamical
  Systems}, vol.~17, no.~1, pp.~909--930, 2018.

\bibitem{arbabi2018data}
H.~Arbabi, M.~Korda, and I.~Mezi{\'c}, ``A data-driven koopman model predictive
  control framework for nonlinear partial differential equations,'' in {\em
  2018 IEEE Conference on Decision and Control (CDC)}, pp.~6409--6414, IEEE,
  2018.

\bibitem{huang2023learning}
D.~Huang, M.~B. Prasetyo, Y.~Yu, and J.~Geng, ``Learning koopman operators with
  control using bi-level optimization,'' in {\em 2023 62nd IEEE Conference on
  Decision and Control (CDC)}, pp.~2147--2152, IEEE, 2023.

\bibitem{boersma2019constrained}
S.~Boersma, B.~Doekemeijer, S.~Siniscalchi-Minna, and J.~van Wingerden, ``A
  constrained wind farm controller providing secondary frequency regulation: An
  les study,'' {\em Renewable energy}, vol.~134, pp.~639--652, 2019.

\end{thebibliography}
\appendices
\section{Autoencoder: Training}\label{Appendix: training}
\begin{figure}[ht]
  \includegraphics[width=0.48\textwidth]{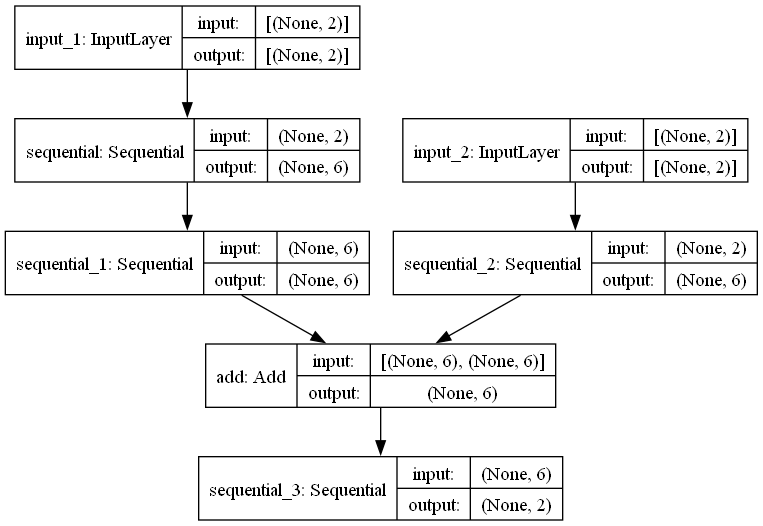}
    \caption{Block diagram NN Architecture for single level training}
    \label{fig:BlockDiagramNNArch1}
\end{figure}

\textbf{Single-level training} In Figure~\ref{fig:BlockDiagramNNArch1}, the depicted block diagram outlines the network architecture utilized to acquire the lifting function vector $g$ and the Koopman operator $K$, simultaneously as described in \eqref{eq:ApproxKalman}. The encoder's activation functions remain nonlinear, while the input consists solely of physical states. Activation functions of \verb|sequential_1|  and \verb|sequential_2| are linear, as are those of \verb|sequential_3| to obtain matrix $C$. This training approach is detailed in algorithm \ref{alg:Single level}. We minimize prediction loss function \eqref{eq:loss_function} to fit the collected data for multiple steps
\begin{align*}
\mathcal{D}=\left\{ \left[
x_{k}^T,u_{k}^T
\right]^T\right\} _{k=1}^{n_{O}-1}.
\end{align*}



\begin{algorithm}
\caption{Single level training}
\label{alg:Single level}

\begin{algorithmic}[1]
\State \textbf{Input:} Training dataset $\mathcal{D}$, learning rate $\eta$, number of epochs $N_E$, batch size $B_t$, number of lifted states $N_g$, number of layers for an encoder, decoder, $A,B$, set linear activation function for all blocks except the encoder and architecture as provided in Figure \ref{fig:BlockDiagramNNArch1}.
\State Initialize weights $w = (A_w,B_w, \mathrm{enc}_w, \mathrm{dec}_w)$  randomly and freeze the biases to $0$. 
\For{$n = 1$ to $N_E$}
    \For{$i = 1$ to $\lceil |\mathcal{D}| / B \rceil$}
        \State Select mini-batch $B_{ti}$ from $D$ with size $B_{t}$
        \State Initialize gradients $\nabla_w \mathcal{L}$ to zero
        \For{each ($X,U, X_+$) in $B_{ti}$}
            \State Perform forward propagation to predict $\hat{X}_+$
            \State Compute loss $\mathcal{L}$ in Eq. \eqref{eq:loss_function}
            \State Backward propagation to compute 
                $\nabla_w \mathcal{L}$
            \State Update gradients using backpropagation
        \EndFor
        \State Update weights using gradients and learning rate
    \EndFor
\EndFor
\State \textbf{Output:} Trained AE Model 1
\end{algorithmic}

\end{algorithm}

\textbf{Bi-level training:} Illustrated in 
 Figure~\ref{fig:BlockDiagramNNArch1}, the network architecture is utilized to acquire the lifting function vector $g$ (\verb|enc|), Koopman operator $K=[A,B]$, and linear decoder (\verb|Cnet|). The encoder's activation functions remain nonlinear, restricted to physical states, while \verb|Cnet|'s activation functions remain linear. We employ Extended Dynamic Mode Decomposition (EDMD) for efficient training. Unlike prior bi-level training approaches, we simultaneously learn lifted states $g$ and $C$ using an AE. Physical states are not included as lifted states, and \verb|Cnet|'s weights are reinitialized using EDMD after the first epoch only. Our goal is to minimize the multi-step prediction loss function \eqref{eq:loss_function} to fit the dataset.

\begin{algorithm}
\caption{Bi-level training}
\label{alg:Bi-level}
\begin{algorithmic}[1]
\State \textbf{Input:} Training dataset $\mathcal{D}$, learning rate $\eta$, number of epochs $N_E$, batch size $B_t$, number of lifted states $N_g$, number of layers for encoder, decoder, set linear activation function for decoder and nonlinear for encoder and Set the Architecture as in fig.\ref{fig:BlockDiagramNNArch2}.
\State Initialize  $\mathrm{enc}_w$ and $\mathrm{dec}_w$  weights randomly and freeze the biases to $0$ 
\For{$n = 1$ to $N_E$}    
    \State Initialize gradients $\nabla_w \mathcal{L}$ to zero
      \State Calculate $G_u,G_+$ as in Eq.\eqref{eq:G_u}
      \State Use EDMD and calculate: $[A,B] = G_{+}G_u^{\dagger}$
        \State Assign $C = \mathrm{dec}_w$
        \State Perform forward propagation to predict $\hat{X}_+$
        \State Compute loss $\mathcal{L}$ using $\hat{X}_+$ and $X_+$
        \State Perform backward propagation to compute $\nabla_w \mathcal{L}$
        \State Update gradients using backpropagation
        \State Update weights using gradients and learning rate  
               \If{$n == 1$}
                \State use EDMD to calculate $C= X_{+}G_+^{\dagger}$
                \State dec.set.weights() = $(C[1,:]+C[2,:])'$
        \EndIf
\EndFor
\State \textbf{Output:}$A,B, C$ and Lifted function as enc
\end{algorithmic}
\end{algorithm}

\begin{align}\label{eq:G_u}
 G_u &=\begin{bmatrix}\mathrm{enc}(x_1) & \cdots& \mathrm{enc}(x_{n_{O}-1})\\
    u_{1} &\cdots& u_{n_{O}-1} \end{bmatrix} \in \mathbb{R}^{(n_g +n_u)\times (n_0-1)}\\
 G_{+} &=\begin{bmatrix}\mathrm{enc}(x_{2}) &\cdots& \mathrm{enc}(x_{n_{O}})
\end{bmatrix} \in \mathbb{R}^{n_g \times (n_0-1)}\\
X_{+} &=\begin{bmatrix}x_{2} &\cdots& x_{n_{O}}
\end{bmatrix} \in \mathbb{R}^{n \times (n_0-1)}
\end{align}

\end{document}